\documentclass[aps,%showpacs,
amsfonts,amssymb,
preprint,
nofootinbib]{revtex4}

\usepackage{natbib}
\usepackage{amsmath}

\usepackage[dvipdfmx]{graphicx}%,hyperref}
\usepackage{amsmath,amssymb,physics,mathtools,comment}
\usepackage{color}
\font\mybb=msbm10 at 12pt

\font\mybbsmall=msbm10 at 10pt

\def\bb#1{\hbox{\mybb#1}}

\def\bbsmall#1{\hbox{\mybbsmall#1}}

\def\ZZsmall {\bbsmall{Z}}

\def\PP {\bb{P}}

\newcommand\beqa{\begin{eqnarray}}
\newcommand\eeqa{\end{eqnarray}}

\begin{document}

{~}

\title{More on Seiberg-Witten Theory and Monstrous Moonshine
}
\vspace{2cm}
\author{Shun'ya Mizoguchi\footnote[1]{E-mail:mizoguch@post.kek.jp},
Takumi Oikawa\footnote[2]{E-mail:oikawat@post.kek.jp},
Hitomi Tashiro\footnote[3]{E-mail:tashiro@post.kek.jp}
and 
Shotaro Yata\footnote[4]{E-mail:yatarou@post.kek.jp}
}

%\vspace{1cm}

\affiliation{\footnotemark[1]Theory Center, 
Institute of Particle and Nuclear Studies,
KEK\\Tsukuba, Ibaraki, 305-0801, Japan 
}

\affiliation{\footnotemark[1]\footnotemark[2]\footnotemark[3]\footnotemark[4]SOKENDAI (The Graduate University for Advanced Studies)\\
Tsukuba, Ibaraki, 305-0801, Japan 
}

\begin{abstract} 
We continue the study of a relationship between the instanton expansion 
of the Seiberg-Witten (SW) prepotential of $D = 4$, ${\cal N }= 2$ $SU(2)$ SUSY 
gauge theory and the Monstrous moonshine. Extending the previous results, 
we show for the cases of $N_f=2$ and $3$ that $q=e^{2\pi i\tau}$, where $\tau$ 
is the complex gauge coupling, again has an expansion whose 
coefficients are all integer-coefficient polynomials 
of the moonshine coefficients of the modular $j$-function 
in terms of an appropriate expansion variable.
We also demonstrate that the new method of calculating the 
SW prepotential developed here is useful by performing some 
explicit computations.
\end{abstract}

\preprint{KEK-TH-2460}
%\pacs{}
\date{November 14, 2022}

\maketitle

\newpage
\title{SWcurve}
\maketitle

%\chapter{}
%%%%%%%%%%%%%%%%%%%%
\section{Introduction}
Recently, an explicit relationship between the instanton expansion 
of the Seiberg-Witten (SW) prepotential \cite{SW1, SW2} 
of $D = 4$, ${\cal N }= 2$ 
$SU(2)$ SUSY gauge theory and the Monstrous moonshine
\cite{McKay,Thompson,MonstrousMoonshine,FLM,FLM2,DGH,Borcherds} 
was studied \cite{SWMoonshine} (See \cite{Tachikawalecture} and \cite{McKayessay} 
for reviews on the respective topics.).
There, by a simple method to obtain SW prepotentials 
using the modular $j$-function, it was shown that 
 the coefficients of the expansion of $q=e^{2\pi i\tau}$ in terms of $A^2\equiv\frac{\Lambda^2}{16a^2}$
$(N_f =0)$ or $\frac{\Lambda^2}{32a^2}$ $(N_f=1)$ 
are all integer-coefficient polynomials of the moonshine coefficients of the 
modular $j$-function\footnote{Here, as in the standard literature on the SW theory, 
$a$ is the eigenvalue of the 
$SU(2)$ adjoint scalar playing the role of the Higgs, and $\tau$ is the complex gauge 
coupling $\frac{\pi i}{g^2}+\frac{\theta}{8\pi}$  identified as the modulus of the 
elliptic fiber over the $u$-plane of the SW curve.}.
In this paper, we continue this study; 
we further examine whether the same phenomenon is also observed 
for $N_f=2$ and $3$ and show that it is indeed the case.

As already emphasized in \cite{SWMoonshine}, 
the relation between the SW theory and the Monstrous moonshine 
is not at all surprising in itself, since SW curves are elliptic curves 
and the modular $j$-function is a bijection between the fundamental domain 
and the complex plane, and thus describes the modulus of an elliptic curve.
On the other hand, it has been known for some time \cite{1207.5139MalmendierOno} 
that the $u$-plane integral \cite{MooreWitten} for a certain ``mock modular form''  
arising in the elliptic genus of K3 yields the $SO(3)$ Donaldson invariants of $\PP^2$ 
(See \cite{Mockingtheuplaneintegral} and references therein for more recent discussions.).
Since the former exhibits a similar moonshine phenomenon called the Mathieu moonshine 
\cite{MathieuMoonshine}, while the latter are certain correlation functions of a  
{\em twisted} 4D ${\cal N}=2$ $SO(3)$ supersymmetric gauge theory \cite{Wittentopological}, 
this observation may be the first example of a connection between a moonshine phenomenon 
and a supersymmetric gauge theory (albeit a twisted one),  which is nontrivial.
In contrast, the correspondence we will use is not that somewhat indirect and intricate, 
but more direct and easy to understand.
Nonetheless, there are other reasons to focus on the relationship.

One of them is as an effective tool for calculating the prepotential. 
This method can be applied to any theory as long as its SW curve is an elliptic curve, 
as in E-string theory. 
We will demonstrate that the method developed in \cite{SWMoonshine}  
for calculating the prepotentials is useful by implementing it in 
$SU(2)$ SW theory with $N_f=2$ and $3$.
To date, 
various ways of computing the SW prepotential are available, but our method is more 
efficient than A-model like computations such as the Nekrasov instanton counting 
\cite{Nekrasov,NekrasovOkounkov}, 
and simpler than other B-model like approaches 
(e.g. \cite{KLT,MasudaSuzuki,Ohta,Ohta2,CDH,DV,EguchiMaruyoshi,IOY,MOT})
as we do not need to 
compute the dual period $a_D$. 
Our approach is simply to develop a series, so there is no way to go wrong 
(as long as one can use computational software like Mathematica). 
For example, by using our method we found 
four typos in the result of the $N_f=3$ prepotential computed by \cite{CDH} long time ago,
as shown in the text \footnote{A similar method for computing the prepotential was already used in 
\cite{ClossetMagureanu2107.03509}. 
The reference to the $j$-function in SW theory itself is of course much older, e.g., the $q$-expansion  for the $j$-function can be found in  \cite{1109.5728} 
(but no reference to the Monstrous moonshine).
}.

Another nontrivial observation in the connection between the SW theory and 
Monstrous moonshine is a possible relationship between two 
different 2d conformal field theories (CFTs)
with different central charges. 
We will show (as was done in \cite{SWMoonshine}) that the quantity $q$ allows an 
expansion in which the coefficients are  {\em integer-coefficient}
polynomials of the moonshine Fourier coefficients. This means that 
$q$ is also expressed in terms of 
integer-coefficient polynomials of the dimensions of representations of the 
Monster group.
As pointed out in \cite{SWMoonshine}, since the dimensions of representations 
of the Monster are related to the 
vertex operator CFT \cite{FLM,FLM2,DGH,Borcherds} 
whereas each instanton contribution of the SW prepotential 
is written as a correlation function of the Liouville theory \cite{AGT}, our observation 
suggests that those are related in some way. 
What exactly that is remains as a problem to be solved in the future.
%
%
%%%%%%%%%%%%%%%%%%%%
\section{SW prepotential from the modular $j$-function}
%%%%%
Let us consider a SW curve given in the Weierstrass form
\begin{equation}
    Y^2=X^3+f(u)X+g(u).\label{eq:eq1}
\end{equation}
We can obtain the 
value of the 
modular $j$-function 
as
\begin{equation}
    j(\tau)=\frac{12^3 \cdot 4f(u)^{3}}{4f(u)^{3}+27g(u)^{2}}.\label{eq:eq2}
\end{equation}
We expand the right hand side 
of this equation \eqref{eq:eq2} by $u$ around $u=\infty$, thereby we get $\frac{1}{u}$-expansion of $\frac{1}{j}$.

On the other hand, the modular $j$-function has an expansion in terms of $q=e^{2\pi i\tau}$ as
\begin{equation}
    j(\tau)=\frac{1}{q}+\sum_{n=0}^{\infty}a_{n}q^{n},\label{eq:eq3}
\end{equation}
whose coefficients $a_{n}$ are integers. Specifically, the first few terms of the expansion \eqref{eq:eq3} are:
\begin{table}[h]
    \centering
    \begin{tabular}{|ccccccc|}
        \hline
        $n$&0&1&2&3&4&5\\
        $a_{n}$&744&196884&21493760&864299970&20245856256&333202640600\\
        \hline
    \end{tabular}
\end{table}\\
Thus we find $\frac{1}{j}$-expansion of $q$ as 
\begin{equation}
   \begin{split}
        q&=\frac{1}{j}+a_{0}\qty(\frac{1}{j})^{2}+\qty(a_{0}^{2}+a_{1})\qty(\frac{1}{j})^{3}+\qty(a_{0}^{3}+3a_{1}a_{0}+a_{2})\qty(\frac{1}{j})^{4}\\
        &\quad+\qty(a_{0}^{4}+6a_{0}^{2}a_{1}+2a_{1}^{2}+4a_{0}a_{2}+a_{3})\qty(\frac{1}{j})^{5}\\
        &\quad +\qty(a_{0}^5+10 a_{0}^3a_{1}+10a_{0}a_{1}^2+10a_{0}^2a_{2}+5a_{1}a_{2}+5a_{0}a_{3}+a_{4})\qty(\frac{1}{j})^6+\order{\qty(\frac{1}{j})^{7}}\label{eq:eq4}
   \end{split}
\end{equation}
by solving inversely the equation \eqref{eq:eq3}.

Substituting $\frac{1}{u}$-expansion of $\frac{1}{j}$ into this equation \eqref{eq:eq4}, we can get $\frac{1}{u}$-expansion of $q$. Furthermore, $\frac{1}{a}$-expansion of $\frac{1}{u}$ is known from the period integral on the SW curve \cite{MasudaSuzuki} and we obtain $\frac{1}{a}$-expansion of $q$. 
%%%%%
Since the complex structure modulus $\tau$ and the SW prepotential $\mathcal{F}$
are related as 
%%%%%
\begin{equation}
    \tau=\pdv[2]{\mathcal{F}}{a},\label{eq:eq5}
\end{equation}
we finally get the expression of the prepotential $\mathcal{F}$ by integrating $\frac{1}{a}$-expansion of $q$ with respect to $a$ twice.

%%%%%
This procedure was performed in \cite{SWMoonshine} 
for $N_f = 0$ and $1$, 
thereby an explicit relationship between the instanton expansion 
and the moonshine was obtained.
Below we will do this for the cases $N_f=2$ and $3$.
%%%%%
%%% v3 %%%
\footnote{It is known \cite{Nahm,MooreWitten} that the monodromy groups 
in these cases are the congruence subgroups $\Gamma(2)$ and 
$\Gamma_0(4)$ of $SL(2,\ZZsmall)$, respectively. The Hauptmoduln for these 
reduced monodromy groups, that is, the weight-0 modular forms associated 
with these congruence subgroups, also allow expansions in term of $q$ 
known as the McKay-Thompson series, whose coefficients are a sum of traces of a 
particular element of the Monster group \cite{MonstrousMoonshine}.  
The analysis of such modular rational elliptic surfaces 
was recently worked out in \cite{Elias2107.04600, Horia2203.03755}.}
%999
%%% v3 %%%

%\section{\texorpdfstring{$N_{f}=2$}{Nf=2} case}
\subsection*{$N_{f}=2$ case}
Let us consider the $\mathcal{N}=2$ $SU(2)$ SUSY gauge theory with $N_{f}=2$ hypermultiplets. The SW curve is given by
\begin{equation}
    y^2=\left(x^2-u+\frac{\Lambda^2}{8}\right)^2+\Lambda^{2}(x+m_{1})(x+m_{2})\label{eq:eq6}
\end{equation}
in the quartic-polynomial representation \cite{MasudaSuzuki}.
This is equivalent to the Weierstrass form
\begin{align}
    Y^{2}&=X^{3}+f(u,m_{1},m_{2})X+g(u,m_{1},m_{2}),\label{eq:eq7}\\
    f(u,m_{1},m_{2})&=-\frac{\Lambda^{4}}{4}+4 \Lambda ^2 {m_1} {m_2}-\frac{16}{3}u^2,\label{eq:eq8}\\
    g(u,m_{1},m_{2})&=\Lambda ^4 \qty(m_1^2+m_2^2)-\frac{2}{3} \qty(\Lambda ^4+8 \Lambda ^2 m_1 m_2)u+\frac{128}{27}u^3.\label{eq:eq9}
\end{align}
Substituting \eqref{eq:eq8} and \eqref{eq:eq9} into the definition of the $j$-function \eqref{eq:eq4}, we have an expansion of $\frac{1}{j}$
\begin{equation}
    \begin{split}
        \frac{1}{j}&=U^2- \qty(\hat{m}_{1}^2+\hat{m}_{2}^2)U^3+\qty[\hat{m}_{1} \hat{m}_{2} \qty(\hat{m}_{1} \hat{m}_{2}+64)-704]U^4\\
        &\quad -72(\hat{m}_{1} \hat{m}_{2}-16) \qty(\hat{m}_{1}^2+\hat{m}_{2}^2)U^5\\
        &\quad  -16 \qty[27 \qty(\hat{m}_{1}^4+\hat{m}_{2}^4)-5 \hat{m}_{1}^3 \hat{m}_{2}^3-102 \hat{m}_{1}^2 \hat{m}_{2}^2+5376 \hat{m}_{1} \hat{m}_{2}-18688]U^6\\
        &\quad -3456 (\hat{m}_{1} \hat{m}_{2}-40) (\hat{m}_{1} \hat{m}_{2}-4) \qty(\hat{m}_{1}^2+\hat{m}_{2}^2)U^7\\
        &\quad -2304 (\hat{m}_{1} \hat{m}_{2}-4) \qty[27 \qty(\hat{m}_{1}^4+\hat{m}_{2}^4)-2 \hat{m}_{1}^3 \hat{m}_{2}^3+30 \hat{m}_{1}^2 \hat{m}_{2}^2+2880 \hat{m}_{1} \hat{m}_{2}-11008]U^8\\
        &\quad -110592 (\hat{m}_{1} \hat{m}_{2}-112) (\hat{m}_{1} \hat{m}_{2}-4)^2 \qty(\hat{m}_{1}^2+\hat{m}_{2}^2)U^9\\
        &\quad -221184 (\hat{m}_{1} \hat{m}_{2}-4)^2 \qty[27 \qty(\hat{m}_{1}^4+\hat{m}_{2}^4)-\hat{m}_{1}^3 \hat{m}_{2}^3+66 \hat{m}_{1}^2 \hat{m}_{2}^2+2112 \hat{m}_{1} \hat{m}_{2}-8576]U^{10}\\
        &\quad +955514880(\hat{m}_{1} \hat{m}_{2}-4)^3 \qty(\hat{m}_{1}^2+\hat{m}_{2}^2)U^{11}\\
        &\quad -1769472 (\hat{m}_{1} \hat{m}_{2}-4)^3 \qty[270 \qty(\hat{m}_{1}^4+\hat{m}_{2}^4)-5 \hat{m}_{1}^3 \hat{m}_{2}^3+816 \hat{m}_{1}^2 \hat{m}_{2}^2+17472 \hat{m}_{1} \hat{m}_{2}-73984]U^{12}\\
        &\quad +382205952 (\hat{m}_{1} \hat{m}_{2}-4)^4 (\hat{m}_{1} \hat{m}_{2}+176) \qty(\hat{m}_{1}^2+\hat{m}_{2}^2)U^{13}\\
        &\quad -254803968 (\hat{m}_{1} \hat{m}_{2}-4)^4 \qty[27 \qty(\hat{m}_{1}^4+\hat{m}_{2}^4)-\hat{m}_{1}^3 \hat{m}_{2}^3+450 \hat{m}_{1}^2 \hat{m}_{2}^2+7680 \hat{m}_{1} \hat{m}_{2}-33536]U^{14}\\
        &\quad +\order{U^{15}}, \label{eq:eq10}
    \end{split}
\end{equation}
where we have introduced $U\equiv\frac{\Lambda^2}{64u},\ \hat{m}_{i}\equiv \frac{8m_{i}}{\Lambda}.$

To relate $a$ and $u$, we use the fact \cite{MasudaSuzuki} that $\pdv{a}{u}$ is given by 
\begin{equation}
    \pdv{a}{u}=\frac{F\qty(\frac{1}{12},\frac{5}{12},1;\frac{12^3}{j})}{\sqrt{2}\qty(-3f(u,m_{1},m_{2}))^{\frac{1}{4}}},\label{eq:eq11}
\end{equation}
where $F(a,b,c;z)$ is a hypergeometric function. Since we have the $U$-expansion of $\frac{1}{j}$ \eqref{eq:eq10}, we can obtain the $U$-expansion of $a$ after substituting \eqref{eq:eq10} into \eqref{eq:eq11} and integrating with respect to $U$. Therefore, defining $A\equiv\frac{\Lambda}{8\sqrt{2}a}$, $U$ is conversely expanded by this $A$ as
\begin{equation}
    \begin{split}
        U&=A^2-8 (\hat{m}_{1} \hat{m}_{2}+1)A^6 +24 \qty(\hat{m}_{1}^2+\hat{m}_{2}^2) A^8 +24\qty(\hat{m}_{1}^2 \hat{m}_{2}^2-8\hat{m}_{1} \hat{m}_{2}+1)A^{10}\\
        &\quad +16 \qty(39+4\hat{m}_{1}\hat{m}_{2})\qty(\hat{m}_{1}^2+\hat{m}_{2}^2) A^{12}\\
        &\quad -8 \qty[81 \qty(\hat{m}_{1}^4+\hat{m}_{2}^4)+56 \hat{m}_{1}^3 \hat{m}_{2}^3+660 \hat{m}_{1}^2 \hat{m}_{2}^2+1128 \hat{m}_{1} \hat{m}_{2}+56]A^{14}+\order{A^{15}}.\label{eq:eq12}
    \end{split}
\end{equation}
By using this expression in \eqref{eq:eq10}, we obtain $A^2$-expansion of $\frac{1}{j}$. Thus, substituting this expansion into \eqref{eq:eq4}, we find $A^2$-expansion of $q$,
\begin{equation}
    \begin{split}
        q&=A^4-\qty(\hat{m}_{1}^2+\hat{m}_{2}^2)A^{6}+\qty[\qty(a_{0}-720)+48\hat{m}_{1}\hat{m}_{2}+\hat{m}_{1}^2\hat{m}_{2}^2]A^{8}\\
        &\quad + \qty[\qty(1224-2a_{0})\qty(\hat{m}_{1}^2+\hat{m}_{2}^2)-48\hat{m}_{1}\hat{m}_{2}\qty(\hat{m}_{1}^2+\hat{m}_{2}^2)]A^{10}\\
        &\quad + \left[\qty(321648-1440a_{0}+a_{0}^2+a_{1})+4(a_{0}-120) \hat{m}_{1}^2 \hat{m}_{2}^2+32 (3 a_{0}-2056) \hat{m}_{1} \hat{m}_{2}\right.\\
        &\quad\quad \left.+48\hat{m}_{1}^3\hat{m}_{2}^3+\qty(a_{0}-504)\qty(\hat{m}_{1}^4+\hat{m}_{2}^4)\right]A^{12}\\
        &\quad +\left[\qty(-666024+3888a_{0}-3a_{0}^2-3a_{1})\qty(\hat{m}_{1}^2+\hat{m}_{2}^2)-\qty(744+2a_{0})\qty(\hat{m}_{1}^2\hat{m}_{2}^4+\hat{m}_{2}^2\hat{m}_{1}^4)\right.\\
        &\quad\quad\left.+\qty(114944-192a_{0})\qty(\hat{m}_{1}\hat{m}_{2}^3+\hat{m}_{2}\hat{m}_{1}^3)\right]A^{14}+\order{A^{16}},\label{eq:eq13}
    \end{split}
\end{equation}
whose coefficients are integer-coefficient polynomials of the moonshine Fourier coefficients $a_{n}$. 

Next, we take logarithm of $q$ as the complex modulus $\tau$ is given by $q=e^{2\pi i \tau}$. 
We have 
\begin{equation}
    \begin{split}
        2\pi i\tau&=\log \qty(A^4)-\qty(\hat{m}_{1}^2+\hat{m}_{2}^2)A^{2}+\frac{1}{2} \qty(2 a_{0}-1440-\hat{m}_{1}^4-\hat{m}_{2}^4+96 \hat{m}_{1} \hat{m}_{2})A^{4}\\
        &\quad -\frac{1}{3} \qty(\hat{m}_{1}^2+\hat{m}_{2}^2) \qty(3 a_{0}-1512+\hat{m}_{1}^4+\hat{m}_{2}^4-\hat{m}_{1}^2 \hat{m}_{2}^2)A^{6}\\
        &\quad +\frac{1}{4} \left[\qty(2 a_{0}^2-2880 a_{0}+4 a_{1}+249792)-\hat{m}_{1}^8-\hat{m}_{2}^8+\qty(4a_{0}+384)\hat{m}_{1}^2 \hat{m}_{2}^2\right.\\
        &\quad\quad \left. +\qty(192a_{0}-124928)\hat{m}_{1}\hat{m}_{2}\right]A^{8}\\
        &\quad -\frac{1}{5}  \left(\hat{m}_{1}^2+\hat{m}_{2}^2\right) \left[\qty(5 a_{0}^2-6120 a_{0}+10 a_{1}-92520)+\hat{m}_{1}^8+\hat{m}_{2}^8\right.\\
        &\quad\quad \left.+\qty(240a_{0}-124800)\hat{m}_{1}\hat{m}_{2}+\hat{m}_{1}^4\hat{m}_{2}^4-\hat{m}_{1}^6\hat{m}_{2}^2-\hat{m}_{1}^2\hat{m}_{2}^6\right]A^{10}\\
        &\quad+\frac{1}{6} \left[-\hat{m}_{1}^{12}-\hat{m}_{2}^{12}+\qty(288a_{0}-87552)\hat{m}_{1}^3\hat{m}_{2}^3\right.\\
        &\quad\quad \left.+\qty(288a_{0}^2-394752a_{0}+576a_{1}+22620672)\hat{m}_{1}\hat{m}_{2}\right.\\
        &\quad\quad \left.+\qty(12 a_{0}^2-2880a_{0}+24a_{1}-7466688)\hat{m}_{1}^2\hat{m}_{2}^2\right.\\
        &\quad\quad \left.+\qty(3a_{0}^2-3024a_{0}+6a_{1}-390096)\qty(\hat{m}_{1}^4+\hat{m}_{2}^4)\right.\\
        &\quad\quad \left.+\qty(2a_{0}^3-4320a_{0}^2+12a_{0}a_{1}+1929888a_{0}-8640a_{1}+6a_{2}-53824512)\right]A^{12}\\
        &\quad+\order{A^{14}}.\label{eq:eq14}
    \end{split}
\end{equation}
Integrating \eqref{eq:eq14} with respect to $a$ twice, we can finally derive $2\pi i$ times the $N_{f}=2$ prepotential. 

Now let us show the results of our calculations. 
The $N_{f}=2$ prepotential $\mathcal{F}^{N_{f}=2}$ takes the form 
\begin{equation}
    \mathcal{F}^{N_{f}=2}=\frac{ia^2}{2\pi}\qty[\log\qty(\frac{a}{\Lambda})^2+(5\log 2-2+i\pi)-\frac{\log a}{2a^2}\sum_{i}^{2}m_{i}^2-\sum_{k=2}^{\infty}\mathcal{F}^{N_f=2}_{k}\qty(\frac{\Lambda}{a})^{2k}],\label{eq:eq15}
\end{equation}
where first few instanton expansion coefficients $\mathcal{F}_{k}^{N_{f}=2}$ is given by
\begin{align}
    \mathcal{F}_{2}^{N_{f}=2}&=\frac{1}{4096}+\frac{1}{32}{\tilde{m}_{1}\tilde{m}_{2}}-\frac{1}{48}\qty(\tilde{m}_{1}^4+\tilde{m}_{2}^4),\label{eq:eq16}\\
    \mathcal{F}_{3}^{N_{f}=2}&=-\frac{3}{8192}\qty(\tilde{m}_{1}^2+\tilde{m}_{2}^2)-\frac{1}{480}\qty(\tilde{m}_{1}^6+\tilde{m}_{2}^6),\label{eq:eq17}\\
    \mathcal{F}_{4}^{N_{f}=2}&=\frac{5}{134217728}+\frac{5}{16384}\tilde{m}_{1}^2\tilde{m}_{2}^2+\frac{5}{196608}\tilde{m}_{1}\tilde{m}_{2}-\frac{1}{2688}\qty(\tilde{m}_{1}^8+\tilde{m}_{2}^8).\label{eq:eq18}
\end{align}
Note that we define $\hat{m}_{i}=\frac{8m_{i}}{\Lambda}=8\tilde{m}_{i}$ and $\mathcal{F}_{k}^{N_{f}=2}$ in the above expression is related to $\mathcal{F}_{k}^{2}$ in \cite{Ohta2} as $\mathcal{F}_{k}^{N_{f}=2}$(here)$=-\frac{2}{\Lambda^{2k}}\mathcal{F}_{k}^{2}$, $\Lambda$(here)$=\Lambda_{2}$.

In this notation, we can obtain the coefficients of the instanton expansion $\mathcal{F}_{k}^{N_{f}=2}$ which include the moonshine Fourier coefficients $a_n$ from the integration of \eqref{eq:eq14}. Here we show  the coefficients of the instanton expansion of $N_{f}=2$ prepotential up to order 4:
\begin{align}
    \mathcal{F}_{2}^{N_{f}=2}&=\frac{1}{(8\sqrt{2})^4\cdot 3\cdot 2}\cdot \frac{1}{2} \left[\qty(2 a_{0}-1440)-\qty(\hat{m}_{1}^{4}+\hat{m}_{2}^4)+96 \hat{m}_{1}\hat{m}_{2}\right]\label{eq:eq22}\\
    \mathcal{F}_{3}^{N_{f}=2}&=-\frac{1}{(8\sqrt{2})^6\cdot 5\cdot 4}\cdot \frac{1}{3} \qty(\hat{m}_{1}^2+\hat{m}_{2}^2) \qty[\qty(3 a_{0}-1512)+\qty(\hat{m}_{1}^4+\hat{m}_{2}^4)-\hat{m}_{1}^2 \hat{m}_{2}^2]\label{eq:eq23}\\
    \mathcal{F}_{4}^{N_{f}=2}&=\frac{1}{(8\sqrt{2})^8\cdot 7\cdot 6}\cdot \frac{1}{4} \left[\qty(2 a_{0}^2-2880a_{0}+4a_{1}+249792)-\qty(\hat{m}_{1}^8+\hat{m}_{2}^8)\right.\notag\\
    &\quad\quad \left. +\qty(4a_{0}+384) \hat{m}_{1}^2 \hat{m}_{2}^2+\qty(192a_{0}-124928) \hat{m}_{1} \hat{m}_{2}\right]\label{eq:eq24}
\end{align}
Substituting the coefficients of $j$-function $a_{n}$ into above expressions, we can derive explicit results of the instanton expansion and we have succeeded to reproduce all known results $\mathcal{F}_{k=2}^{N_{f}=2}$ for $k=2, 3, 4$ from the coefficients of $A^4, A^6, A^8$ in \eqref{eq:eq14}, respectably.

Indeed, substituting $a_0=744$ and $\hat{m}_{i}=8\tilde{m}_{i}$ into \eqref{eq:eq22}, then we get
\begin{align}
    \mathcal{F}_{2}^{N_{f}=2}&=\dfrac{\frac{1}{2} \left[\qty(2 a_{0}-1440)-8^4\qty(\tilde{m}_{1}^{4}+\tilde{m}_{2}^4)+96\cdot 8^2 \tilde{m}_{1}\tilde{m}_{2}\right]}{(8\sqrt{2})^4\cdot 3\cdot 2}\notag\\
    &=\qty[\frac{1}{4096}+\frac{1}{32}{\tilde{m}_{1}\tilde{m}_{2}}-\frac{1}{48}\qty(\tilde{m}_{1}^4+\tilde{m}_{2}^4)]\notag
\end{align}
which is consistent with the known $k=2$ expression of $\mathcal{F}^{N_{f}=2}$ in \cite{Ohta2}. 
Also, the expansion coefficients $\mathcal{F}_{k}^{N_{f}=2}$ for larger $k$ 
can similarly be calculated from the coefficients of $A^{2k}$ in \eqref{eq:eq14}.

%\section{\texorpdfstring{$N_{f}=3$}{Nf=3} case}
\subsection*{$N_{f}=3$ case}

A similar result can be obtained in the case of $N_{f}=3$. 
The SW curve in this case is 
\begin{equation}
    y^2=\qty[x^2-u+\frac{\Lambda}{4}\qty(x+\frac{m_{1}+m_{2}+m_{3}}{2})]^2+\Lambda(x+m_{1})(x+m_{2})(x+m_{3})\label{eq:eq26}
\end{equation}
in the quartic-polynomial representation \cite{MasudaSuzuki}. This is equivalent to the Weierstrass form
\begin{align}
    Y^{2}&=X^{3}+f(u,m_{1},m_{2},m_{3})X+g(u,m_{1},m_{2},m_{3}),\label{eq:eq27}\\
    f(u,m_{1},m_{2},m_{3})&=-\frac{\Lambda ^4}{768}-\frac{\Lambda^{2}}{4} \qty(m_{1}^2+m_{2}^2+m_{3}^2)+4 \Lambda  m_{1}m_{2}m_{3}-\frac{16 u^2}{3}+\frac{\Lambda ^2 u}{3},\label{eq:eq28}\\
    g(u,m_{1},m_{2},m_{3})&=\frac{\Lambda ^6}{55296}+\frac{\Lambda^4}{192} \qty(m_{1}^2+m_{2}^2+m_{3}^2)+\Lambda ^2 \qty(m_{1}^{2}m_{2}^{2}+m_{2}^{2}m_{3}^{2}+m_{3}^{2}m_{1}^{2}) \notag\\
    &\quad -u \qty[\frac{\Lambda ^4}{144}+\frac{2}{3} \Lambda ^2 \qty(m_{1}^2+m_{2}^2+m_{3}^2)+\frac{16}{3} \Lambda  m_{1}m_{2}m_{3}]\notag\\
    &\quad -\frac{\Lambda^3}{12} m_{1}m_{2}m_{3} +\frac{128}{27}u^3+\frac{5 \Lambda ^2}{9} u^{2}.\label{eq:eq29}
\end{align}
From the definition of the $j$-function \eqref{eq:eq2} and then using \eqref{eq:eq28} and \eqref{eq:eq29}, we get an expansion of $\frac{1}{j}$,
\begin{equation}
    \begin{split}
        \frac{1}{j}&=-U+ \qty(\hat{m}_{1}^2+\hat{m}_{2}^2+\hat{m}_{3}^2-752)U^2\\
        &\quad + \qty[1472\qty(\hat{m}_{1}^2+\hat{m}_{2}^2+\hat{m}_{3}^2)-\qty(\hat{m}_{1}^2 \hat{m}_{2}^2+\hat{m}_{2}^2\hat{m}_{3}^2+\hat{m}_{3}^2\hat{m}_{1}^2)-56\hat{m}_{1}\hat{m}_{2}\hat{m}_{3}-368640]U^{3}\\
        &\quad +\left[-704\qty(\hat{m}_{1}^4+\hat{m}_{2}^4+\hat{m}_{3}^4)-2656\qty(\hat{m}_{1}^2\hat{m}_{2}^2+\hat{m}_{2}^2\hat{m}_{3}^2+\hat{m}_{3}^2\hat{m}_{1}^2)+\hat{m}_{1}^2\hat{m}_{2}^2\hat{m}_{3}^2\right.\\
        &\quad\quad \left. +64 \hat{m}_{1} \hat{m}_{2} \hat{m}_{3} \qty(\hat{m}_{1}^2+\hat{m}_{2}^2+\hat{m}_{3}^2)+1057792\qty(\hat{m}_{1}^2+\hat{m}_{2}^2+\hat{m}_{3}^2)-149094400\right]U^{4}\\
        &\quad + 8\left[-123392\qty(\hat{m}_{1}^4+\hat{m}_{2}^4+\hat{m}_{3}^4)+70123520\qty(\hat{m}_{1}^2+\hat{m}_{2}^2+\hat{m}_{3}^2)-7548928\hat{m}_{1}\hat{m}_{2}\hat{m}_{3}\right.\\
        &\quad\quad \left. +144\qty(\hat{m}_{1}^4\hat{m}_{2}^2+\hat{m}_{2}^4\hat{m}_{3}^2+\hat{m}_{3}^4\hat{m}_{1}^2)-337408\qty(\hat{m}_{1}^2\hat{m}_{2}^2+\hat{m}_{2}^2\hat{m}_{3}^2+\hat{m}_{3}^2\hat{m}_{1}^2)\right.\\
        &\quad\quad \left. +21376\hat{m}_{1}\hat{m}_{2}\hat{m}_{3}\qty(\hat{m}_{1}^2+\hat{m}_{2}^2+\hat{m}_{3}^2)-9\hat{m}_{1}\hat{m}_{2}\hat{m}_{3}\qty(\hat{m}_{1}^2\hat{m}_{2}^2+\hat{m}_{2}^2\hat{m}_{3}^2+\hat{m}_{3}^2\hat{m}_{1}^2)\right.\\
        &\quad\quad \left. -6748635136\right]U^5\\
        &\quad + 16\left[18688\qty(\hat{m}_{1}^6+\hat{m}_{2}^6+\hat{m}_{3}^6)-48361472\qty(\hat{m}_{1}^4+\hat{m}_{2}^4+\hat{m}_{3}^4)+5\hat{m}_{1}^2\hat{m}_{2}^2\hat{m}_{3}^2\right.\\
        &\quad\quad \left. +15678832640\qty(\hat{m}_{1}^2+\hat{m}_{2}^2+\hat{m}_{3}^2)+116686848\qty(\hat{m}_{1}^2\hat{m}_{2}^2+\hat{m}_{2}^2\hat{m}_{3}^2+\hat{m}_{3}^2\hat{m}_{1}^2)\right.\\
        &\quad\quad \left.+136704\qty(\hat{m}_{1}^4\hat{m}_{2}^2+\hat{m}_{2}^4\hat{m}_{3}^2+\hat{m}_{3}^4\hat{m}_{1}^2)-27\qty(\hat{m}_{1}^4\hat{m}_{2}^4+\hat{m}_{2}^4\hat{m}_{3}^4+\hat{m}_{3}^4\hat{m}_{1}^4)\right.\\
        &\quad\quad \left. -2016542720\hat{m}_{1}\hat{m}_{2}\hat{m}_{3}+11020288\hat{m}_{1}\hat{m}_{2}\hat{m}_{3}\qty(\hat{m}_{1}^2+\hat{m}_{2}^2+\hat{m}_{3}^2)\right.\\
        &\quad\quad\left. -5376\hat{m}_{1}\hat{m}_{2}\hat{m}_{3}\qty(\hat{m}_{1}^4+\hat{m}_{2}^4+\hat{m}_{3}^4)-20544\hat{m}_{1}\hat{m}_{2}\hat{m}_{3}\qty(\hat{m}_{1}^2\hat{m}_{2}^2+\hat{m}_{2}^2\hat{m}_{3}^2+\hat{m}_{3}^2\hat{m}_{1}^2)\right.\\
        &\quad\quad \left. +102\hat{m}_{1}^2\hat{m}_{2}^2\hat{m}_{3}^2\qty(\hat{m}_{1}^2+\hat{m}_{2}^2+\hat{m}_{3}^2)-1136991928320\right]U^6+\order{U^7}\label{eq:eq30}
    \end{split}
\end{equation}
where we have introduced $U\equiv\frac{\Lambda^2}{4096u},\ \hat{m}_{i}\equiv \frac{64m_{i}}{\Lambda}.$

Similarly to the $N_{f}=2$ case, we again use \eqref{eq:eq11} and then get the $U$-expansion of $a$. Defining $A\equiv\frac{\Lambda}{64\sqrt{2}a}$, 
$U$ is conversely expanded by this $A$ as
\begin{equation}
    \begin{split}
        U&=A^2-8 A^4-8 \qty[\left(\hat{m}_{1}^2+\hat{m}_{2}^2+\hat{m}_{3}^2\right)+\hat{m}_{1} \hat{m}_{2} \hat{m}_{3}-7]A^{6}\\
        &\quad +8 \qty[22\qty(\hat{m}_{1}^2+\hat{m}_{2}^2+\hat{m}_{3}^2)+3\qty(\hat{m}_{1}^2\hat{m}_{2}^2+\hat{m}_{2}^2\hat{m}_{3}^2+\hat{m}_{3}^2\hat{m}_{1}^2)+40\hat{m}_{1}\hat{m}_{2}\hat{m}_{3}-48]A^{8}\\
        &\quad +8\left[3\qty(\hat{m}_{1}^4+\hat{m}_{2}^4+\hat{m}_{3}^4)-312\qty(\hat{m}_{1}^2+\hat{m}_{2}^2+\hat{m}_{3}^2)-132\qty(\hat{m}_{1}^2\hat{m}_{2}^2+\hat{m}_{2}^2\hat{m}_{3}^2+\hat{m}_{3}^2\hat{m}_{1}^2)\right.\\
        &\quad \quad\left.-840\hat{m}_{1}\hat{m}_{2}\hat{m}_{3}-24\hat{m}_{1}\hat{m}_{2}\hat{m}_{3}\qty(\hat{m}_{1}^2+\hat{m}_{2}^2+\hat{m}_{3}^2)+3\hat{m}_{1}^2\hat{m}_{2}^2\hat{m}_{3}^2+323\right]A^{10}\\
        &\quad +16 \left[-48\qty(\hat{m}_{1}^4+\hat{m}_{2}^4+\hat{m}_{3}^4)+1806\qty(\hat{m}_{1}^2+\hat{m}_{2}^2+\hat{m}_{3}^2)+1616\qty(\hat{m}_{1}^2\hat{m}_{2}^2+\hat{m}_{2}^2\hat{m}_{3}^2+\hat{m}_{3}^2\hat{m}_{1}^2)\right.\\
        &\quad\quad \left. +39\qty(\hat{m}_{1}^4\hat{m}_{2}^2+\hat{m}_{2}^4\hat{m}_{3}^2+\hat{m}_{3}^4\hat{m}_{1}^2)+6528\hat{m}_{1}\hat{m}_{2}\hat{m}_{3}+728\hat{m}_{1}\hat{m}_{2}\hat{m}_{3}\qty(\hat{m}_{1}^2+\hat{m}_{2}^2+\hat{m}_{3}^2)\right.\\
        &\quad\quad \left.+4\hat{m}_{1}\hat{m}_{2}\hat{m}_{3}\qty(\hat{m}_{1}^2\hat{m}_{2}^2+\hat{m}_{2}^2\hat{m}_{3}^2+\hat{m}_{3}^2\hat{m}_{1}^2)+268\hat{m}_{1}^2\hat{m}_{2}^2\hat{m}_{3}^2-1080\right]A^{12}\\
        &\quad -8 \left[56\qty(\hat{m}_{1}^6+\hat{m}_{2}^6+\hat{m}_{3}^6)-1802\qty(\hat{m}_{1}^4+\hat{m}_{2}^4+\hat{m}_{3}^4)+37160\qty(\hat{m}_{1}^2+\hat{m}_{2}^2+\hat{m}_{3}^2)\right.\\
        &\quad\quad \left. +58704\qty(\hat{m}_{1}^2\hat{m}_{2}^2+\hat{m}_{2}^2\hat{m}_{3}^2+\hat{m}_{3}^2\hat{m}_{1}^2)+5496\qty(\hat{m}_{1}^4\hat{m}_{2}^2+\hat{m}_{2}^4\hat{m}_{3}^2+\hat{m}_{3}^4\hat{m}_{1}^2)+56\hat{m}_{1}^2\hat{m}_{2}^2\hat{m}_{3}^2\right.\\
        &\quad\quad\left. +81\qty(\hat{m}_{1}^4\hat{m}_{2}^4+\hat{m}_{2}^4\hat{m}_{3}^4+\hat{m}_{3}^4\hat{m}_{1}^4)+169928\hat{m}_{1}\hat{m}_{2}\hat{m}_{3}+45504\hat{m}_{1}\hat{m}_{2}\hat{m}_{3}\qty(\hat{m}_{1}^2+\hat{m}_{2}^2+\hat{m}_{3}^2)\right.\\
        &\quad\quad\left. +1128\hat{m}_{1}\hat{m}_{2}\hat{m}_{3}\qty(\hat{m}_{1}^4+\hat{m}_{2}^4+\hat{m}_{3}^4)+6176\hat{m}_{1}\hat{m}_{2}\hat{m}_{3}\qty(\hat{m}_{1}^2\hat{m}_{2}^2+\hat{m}_{2}^2\hat{m}_{3}^2+\hat{m}_{3}^2\hat{m}_{1}^2)\right.\\
        &\quad \quad\left. +41832\hat{m}_{1}^2\hat{m}_{2}^2\hat{m}_{3}+660\hat{m}_{1}^2\hat{m}_{2}^2\hat{m}_{3}^2\qty(\hat{m}_{1}^2+\hat{m}_{2}^2+\hat{m}_{3}^2)-14344\right]A^{14}+\order{A^{16}}.\label{eq:eq31}
    \end{split}
\end{equation}
By using this expression in \eqref{eq:eq30}, we obtain $A^2$-expansion of $\frac{1}{j}$. Thus, substituting this expansion into \eqref{eq:eq4} and taking logarithm, we find $A^2$-expansion of $q$,
\begin{equation}
    \begin{split}
        q&=-A^2+\qty[\qty(a_{0}-744)+\qty(\hat{m}_{1}^2+\hat{m}_{2}^2+\hat{m}_{3}^2)]A^4\\
        &\quad +\left[\qty(-356664+1488a_{0}-a_{0}^2-a_{1})+\qty(1464-2a_{0})\qty(\hat{m}_{1}^2+\hat{m}_{2}^2+\hat{m}_{3}^2)\right.\\
        &\quad\quad \left.-\qty(\hat{m}_{1}^2\hat{m}_{2}^2+\hat{m}_{2}^2\hat{m}_{3}^2+\hat{m}_{3}^2\hat{m}_{1}^2)-48\hat{m}_{1}\hat{m}_{2}\hat{m}_{3}\right]A^6\\
        &\quad \left[\qty(-140379008+1266864a_{0}-2232a_{0}^2+a_{0}^3-2232a_{1}+3a_{0}a_{1}+a_{2})\right.\\
        &\quad\quad\left.+\qty(1034496-4416a_{0}+3a_{0}^2+3a_{1})\qty(\hat{m}_{1}^2+\hat{m}_{2}^2+\hat{m}_{3}^2)+\hat{m}_{1}^2\hat{m}_{2}^2\hat{m}_{3}^2\right.\\
        &\quad\quad \left.+\qty(a_{0}-720)\qty(\hat{m}_{1}^4+\hat{m}_{2}^4+\hat{m}_{3}^4)+\qty(4a_{0}-2688)\qty(\hat{m}_{1}^2\hat{m}_{2}^2+\hat{m}_{2}^2\hat{m}_{3}^2+\hat{m}_{3}^2\hat{m}_{1}^2)\right.\\
        &\quad\quad \left.+\qty(96a_{0}-70144)\hat{m}_{1}\hat{m}_{2}\hat{m}_{3}+48\hat{m}_{1}\hat{m}_{2}\hat{m}_{3}\qty(\hat{m}_{1}^2+\hat{m}_{2}^2+\hat{m}_{3}^2)\right]A^{8}+\order{A^{10}}.
    \end{split}
\end{equation}
Its logarithm yields
\begin{equation}
    \begin{split}
        2\pi i\tau &=\log \left(-A^2\right)+ \left[\qty(744-a_{0})-\qty(\hat{m}_{1}^2+\hat{m}_{2}^2+\hat{m}_{3}^2)\right]A^2\\
        &\quad +\frac{1}{2} \left[\qty(159792-1488a_{0}+a_{0}^2+2a_{1})+\qty(2a_{0}-1440)\qty(\hat{m}_{1}^2+\hat{m}_{2}^2+\hat{m}_{3}^2)\right.\\
        &\quad \quad \left.-\qty(\hat{m}_{1}^4+\hat{m}_{2}^4+\hat{m}_{3}^4)+96\hat{m}_{1}\hat{m}_{2}\hat{m}_{3}\right]A^4\\
        &\quad +\frac{1}{3}  \left[\qty(36893760-1069992a_{0}+2232a_{0}^2-a_{0}^3+4464a_{1}-6a_{0}a_{1}-3a_{2})\right.\\
        &\quad \quad \left. +\qty(4392a_{0}-3a_{0}^2-6a_{1}-426456)\qty(\hat{m}_{1}^2+\hat{m}_{2}^2+\hat{m}_{3}^2)-\qty(\hat{m}_{1}^6+\hat{m}_{2}^6+\hat{m}_{3}^6)\right.\\
        &\quad\quad\left.+\qty(1512-3a_{0})\qty(\hat{m}_{1}^2\hat{m}_{2}^2+\hat{m}_{2}^2\hat{m}_{3}^2+\hat{m}_{3}^2\hat{m}_{1}^2)+\qty(103296-144a_{0})\hat{m}_{1}\hat{m}_{2}\hat{m}_{3}\right]A^6\\
        &\quad +\frac{1}{4}\left[\left(8515094496-561516032a_{0}+2533728a_{0}^2-2976a_{0}^3+a_{0}^4+5067456a_{1}\right.\right.\\
        &\quad\quad \left.\left.-17856a_{0}a_{1}+12a_{0}^2 a_{1}+6a_{1}^2-8928a_{2}+12a_{0}a_{2}+4a_{3}\right)+4a_{0}\hat{m}_{1}^2\hat{m}_{2}^2\hat{m}_{3}^2\right.\\
        &\quad\quad \left.+\left(4137984a_{0}-8832a_{0}^2+4a_{0}^3-17664a_{1}+24a_{0}a_{1}+12a_{2}\right.\right.\\
        &\quad\quad\left.\left.-132871168\right)\qty(\hat{m}_{1}^2+\hat{m}_{2}^2+\hat{m}_{3}^2)-\qty(\hat{m}_{1}^8+\hat{m}_{2}^8+\hat{m}_{3}^8)\right.\\
        &\quad\quad \left.+\qty(249792-2880a_{0}+2a_{0}^2+4a_{1})\qty(\hat{m}_{1}^4+\hat{m}_{2}^4+\hat{m}_{3}^4)\right.\\
        &\quad\quad +\left.\qty(26941440-280576a_{0}+192a_{0}^2+384a_{1})\hat{m}_{1}\hat{m}_{2}\hat{m}_{3}\right.\\
        &\quad\quad\left.+\qty(454656-10752a_{0}+8a_{0}^2+16a_{1})\qty(\hat{m}_{1}^2\hat{m}_{2}^2+\hat{m}_{2}^2\hat{m}_{3}^2+\hat{m}_{3}^2\hat{m}_{1}^2)\right.\\
        &\quad\quad\left.+\qty(192a_{0}-124928)\hat{m}_{1}\hat{m}_{2}\hat{m}_{3}\qty(\hat{m}_{1}^2+\hat{m}_{2}^2+\hat{m}_{3}^2)\right]A^8+\order{A^{10}}.\label{eq:eq32}
    \end{split}
\end{equation}
Integrating \eqref{eq:eq32} with respect to $a$ twice, we can finally derive $2\pi i$ times the $N_{f}=3$ prepotential. 

In this case, the $N_{f}=3$ prepotential $\mathcal{F}^{N_{f}=3}$ is
\begin{equation}
    \mathcal{F}^{N_{f}=3}=\frac{ia^2}{2\pi}\qty[\frac{1}{2}\log\qty(\frac{a}{\Lambda})^2+\frac{1}{2}(9\log 2-2-i\pi)-\frac{\log a}{2a^2}\sum_{i}^{2}m_{i}^2-\sum_{k=2}^{\infty}\mathcal{F}^{N_f=3}_{k}\qty(\frac{\Lambda}{a})^{2k}],\label{eq:eq33}
\end{equation}
where first few instanton expansion coefficients $\mathcal{F}_{k}^{N_{f}=3}$ is given in \cite{Ohta2} by
\begin{align}
    \mathcal{F}_{2}^{N_{f}=3}&=\frac{1}{33554432}+\frac{1}{4096}\left({\tilde{m}_{1}^2+\tilde{m}_{2}^2+\tilde{m}_{3}^2}\right)+\frac{1}{32}{\tilde{m}_{1}\tilde{m}_{2}\tilde{m}_{3}}+\frac{1}{48}\left({\tilde{m}_{1}^4+\tilde{m}_{2}^4+\tilde{m}_{3}^4}\right),\label{eq:eq34}\\
    \mathcal{F}_{3}^{N_{f}=3}&=-\frac{3}{33554432}\left(\tilde{m}_{1}^2+\tilde{m}_{2}^2+\tilde{m}_{3}^2\right)-\frac{1}{32768}\tilde{m}_{1}\tilde{m}_{2}\tilde{m}_{3}\notag\\
    &\quad -\frac{3}{8192}\left(\tilde{m}_{1}^2\tilde{m}_{2}^2+\tilde{m}_{2}^2\tilde{m}_{3}^2+\tilde{m}_{3}^2\tilde{m}_{1}^2\right)-\frac{1}{480}\left(\tilde{m}_{1}^6+\tilde{m}_{2}^6+\tilde{m}_{3}^6\right)\label{eq:eq35}\\
    \mathcal{F}_{4}^{N_{f}=3}&=\frac{5}{4503599627370496}+\frac{5}{103079215104}\left(\tilde{m}_{1}^2+\tilde{m}_{2}^2+\tilde{m}_{3}^2\right)+\frac{7}{268435456}\tilde{m}_{1}\tilde{m}_{2}\tilde{m}_{3}\notag\\
    &\quad +\frac{5}{134217728}\left(\tilde{m}_{1}^4+\tilde{m}_{2}^4+\tilde{m}_{3}^4\right)+\frac{25}{33554432}\left(\tilde{m}_{1}^2\tilde{m}_{2}^2+\tilde{m}_{2}^2\tilde{m}_{3}^2+\tilde{m}_{3}^2\tilde{m}_{1}^2\right)\notag\\
    &\quad +\frac{5}{196608}\tilde{m}_{1}\tilde{m}_{2}\tilde{m}_{3}\left(\tilde{m}_{1}^2+\tilde{m}_{2}^2+\tilde{m}_{3}^2\right)+\frac{5}{16384}\tilde{m}_{1}^{2}\tilde{m}_{2}^{2}\tilde{m}_{3}^{2}-\frac{1}{2688}\left(\tilde{m}_{1}^{8}+\tilde{m}_{2}^{8}+\tilde{m}_{3}^{8}\right).\label{eq:eq36}
\end{align}
Note that we define $\hat{m}_{i}=\frac{64m_{i}}{\Lambda}=64\tilde{m}_{i}$ and $\mathcal{F}_{k}^{N_{f}=3}$ in the above expression is related to $\mathcal{F}_{k}^{3}$ in \cite{Ohta2} as $\mathcal{F}_{k}^{N_{f}=3}$(here)$=-\frac{2}{\Lambda^{2k}}\mathcal{F}_{k}^{3}$, $\Lambda$(here)$=\Lambda_{3}$.

In this notation, we have succeeded to reproduce all known results $\mathcal{F}_{k=2}^{N_{f}=3}$ for $k=2, 3, 4$ from the coefficients of $A^4, A^6, A^8$ in \cite{Ohta2}, respectively.

For example, we take the third term of \eqref{eq:eq32} and integrating with respect to $a$ twice, we get the coefficient of $A^4$ term as
\begin{align}
    \mathcal{F}_{2}^{N_{f}=3}&=\frac{1}{(64\sqrt{2})^4\cdot 3\cdot 2}\cdot \frac{1}{2} \left[\qty(159792-1488a_{0}+a_{0}^2+2a_{1})+\qty(2a_{0}-1440)\qty(\hat{m}_{1}^2+\hat{m}_{2}^2+\hat{m}_{3}^2)\right.\notag\\
    &\quad \quad \left.-\qty(\hat{m}_{1}^4+\hat{m}_{2}^4+\hat{m}_{3}^4)+96\hat{m}_{1}\hat{m}_{2}\hat{m}_{3}\right]\notag\\
    &=\frac{1}{33554432}+\frac{1}{4096}\left({\tilde{m}_{1}^2+\tilde{m}_{2}^2+\tilde{m}_{3}^2}\right)+\frac{1}{32}{\tilde{m}_{1}\tilde{m}_{2}\tilde{m}_{3}}+\frac{1}{48}\left({\tilde{m}_{1}^4+\tilde{m}_{2}^4+\tilde{m}_{3}^4}\right)\notag
\end{align}
which corresponds to $\mathcal{F}_{2}^{N_{f}=3}$ in \eqref{eq:eq34} correctly. 

\section{Comparison with Nekrasov Partition Function}
%%%%%
As is well known, the SW prepotential is also obtained by Instanton counting by Nekrasov
\cite{Nekrasov,NekrasovOkounkov}. 
Our results are of course in full agreement with those obtained that way. 
We will see this briefly below.
%%%%%

\subsection*{$N_f=3$}
The instanton partition function is calculated as a sum over all possible Young tableaus parametrized as $Y=(\lambda_1 \geq \lambda_2\geq \cdots )$, 
where $\lambda_{\ell}$ is the height of the $\ell$-th column.
In the case of $D = 4$, ${\cal N }= 2$ $SU(2)$ gauge theory with $N_{f}=3$, the instanton partition function is \footnote{We use the notation used in \cite{EguchiMaruyoshi}, where 
the terms $\epsilon_1L_{Y_j}(t)$ and $-\epsilon_2A_{Y_i}(t)$ 
in the second line of eq.(A.3) in \cite{EguchiMaruyoshi}  should read 
$\epsilon_1L_{Y_i}(t)$ and $-\epsilon_2A_{Y_j}(t)$, respectively, 
and are corrected in (\ref{eq:eq39}).
}
\begin{equation}
   Z^{N_f=3}_{\text{inst}}=\sum_{(Y_1,Y_2)}{\Lambda^{|\vec{Y}|}_3Z_{\text{vector}}(\vec{a},\vec{Y})
   Z_{\text{antifund}}(\vec{a},\vec{Y},\mu_1)Z_{\text{antifund}}(\vec{a},\vec{Y},\mu_2)Z_{\text{fund}}(\vec{a},\vec{Y},-\mu_3)}.
   \label{eq:eq38}
\end{equation}
Here
\begin{equation}
   \begin{split}
      Z_{\text{vector}}(a,\vec{Y})=&\prod_{i,j=1,2}\prod_{s\in Y_i}(2a\delta_{ij}-\epsilon_1L_{Y_j}(s)+\epsilon_2(A_{Y_i}(s)+1))^{-1}
      \\&\quad\quad\times\prod_{t\in Y_j}(-2a\delta_{ij}+\epsilon_1L_{Y_i}(t)-\epsilon_2(A_{Y_j}(t)+1)+\epsilon_{+})^{-1},
      \\Z_{\text{fund}}(a,\vec{Y},\mu)=&
      \prod_{i=1,2}\prod_{s\in Y_i}(a\delta_i+\epsilon_1(l-1)+\epsilon_2(m-1)-\mu+\epsilon_+),
      \\Z_{\text{antifund}}(a,\vec{Y},\mu)=&
      \prod_{i=1,2}\prod_{s\in Y_i}(a\delta_i+\epsilon_1(l-1)+\epsilon_2(m-1)+\mu),
      \label{eq:eq39}
   \end{split}
\end{equation}
where we define $\epsilon_+=\epsilon_1+\epsilon_2$,\ $\delta_1=+1$,\ $\delta_2=-1$ and 
\begin{equation}
   \delta_{ij}=\left\{
      \begin{aligned}
         0\ &\text{for}\ i=j,\\
         1\ &\text{for}\ i=1\ \text{and}\ j=2,\\
         -1\ &\text{for}\ i=2\ \text{and}\ j=1.
      \end{aligned}
   \right.
   \label{eq:eq40}
\end{equation}
For a box $s$ at the coordinate $(\ell, m)$, the leg-length $L_Y(s)=\lambda'_m -\ell$ and the arm-length $A_Y(s)=\lambda_{\ell}-m$ where $\lambda'_m$ is the length of the $m$-th row.
The prepotential can be obtained in the limit where the deformation 
parameters go to zero (with a fixed ratio $\epsilon_1/\epsilon_2$):
\begin{equation}
   \mathcal{F}_{\text{inst}}=\lim_{\epsilon_{1,2}\rightarrow 0}(-\epsilon_1 \epsilon_2)\log Z_{\text{inst}}.
   \label{eq:eq41}
\end{equation}
Therefore, $F^{N_f=3}$ is expanded as
\begin{equation}
   \begin{split}
      \mathcal{F}^{N_{f}=3}=&
      \left(\frac{1}{33554432}+\frac{\tilde{m}_1^2+\tilde{m}_2^2+\tilde{m}_3^2}{4096}
      +\frac{\tilde{m}_1\tilde{m}_2\tilde{m}_3}{32}\right)\frac{\Lambda^4}{a^2}
      \\&\quad
      -\left(\frac{3}{33554432}(\tilde{m}_1^2+\tilde{m}_2^2+\tilde{m}_3^2)
      +\frac{\tilde{m}_1\tilde{m}_2\tilde{m}_3}{32768}
      +\frac{3}{8192}\left(\tilde{m}_1^2\tilde{m}_2^2+\tilde{m}_2^2\tilde{m}_3^2
      +\tilde{m}^2_3\tilde{m}^2_1\right)\right)\frac{\Lambda^6}{a^4}
      \\&\quad
      +\left(\frac{5}{4503599627370496}
      +\frac{5}{134217728}(\tilde{m}^2_1+\tilde{m}^2_2+\tilde{m}^2_3)
      +\frac{7}{268435456}\tilde{m}_1\tilde{m}_2\tilde{m}_3\right.
      \\&\quad
      +\frac{5}{134217728}(\tilde{m}^4_1+\tilde{m}^4_2+\tilde{m}^4_3)
      +\frac{25}{33554432}(\tilde{m}_1^2\tilde{m}_2^2+\tilde{m}_2^2\tilde{m}_3^2+\tilde{m}_3^2\tilde{m}_1^2)
      \\&\quad
      +\frac{5}{196608}\tilde{m}_1\tilde{m}_2\tilde{m}_3(\tilde{m}_1^2+\tilde{m}_2^2+\tilde{m}_3^2)
      \left.
      +\frac{5}{16384}\tilde{m}^2_1\tilde{m}^2_2\tilde{m}_3^2\right)\frac{\Lambda^8 }{a^6}+\mathcal{O}(\Lambda^9).
   \end{split}
   \label{eq:eq42}
\end{equation}
Note that we have defined $\tilde{m}_i\equiv\frac{\Lambda}{\sqrt{2}}\mu_i$ $(i=1,2,3)$ and $\Lambda=\frac{\Lambda_3}{4\sqrt{2}}$.

\subsection*{$N_f=2$}

Next, let us consider the case of $N_f=2$. We take a limit where $\mu_2 \rightarrow \infty$ while keeping $\mu_2\Lambda_3\equiv \Lambda^2_2$ fixed.
In this limit, the partition function becomes:
\begin{equation}
   Z^{N_f=2}_{\text{inst}} = \sum_{(Y_1,Y_2)}\Lambda^{2|\vec{Y}|}_2 Z_{\text{vector}}(a,\vec{Y})Z_{\text{antifund}}(a,\vec{Y},\mu_1)Z_{\text{fund}}(a,\vec{Y},-\mu_3).
   \label{eq:eq43}
\end{equation}
Therefore, the expansion of $\mathcal{F}^{N_f=2}$ is
\begin{equation}
   \begin{split}
      \mathcal{F}^{N_f=2}=&
      \frac{1}{16}\Lambda^2+\left(\frac{1}{4096}+\frac{1}{32}\tilde{m}_1 \tilde{m}_2\right)\frac{\Lambda ^4}{a^2}
      -\frac{3}{8192}\left(\tilde{m}_1^2+\tilde{m}_2^2\right)\frac{\Lambda^6}{a^4}
      \\&\quad
      +\left(\frac{5}{134217728}+  \frac{5}{196608}\tilde{m}_1 \tilde{m}_2 + \frac{5}{16384}\tilde{m}^2_1 \tilde{m}^2_2\right) \frac{\Lambda ^8}{a^6}+\mathcal{O}(\Lambda^9),
   \end{split}
   \label{eq:eq44}
\end{equation}
where $\tilde{m}_1\equiv\frac{\Lambda}{\sqrt{2}}\mu_1$, $\tilde{m}_2\equiv\frac{\Lambda}{\sqrt{2}}\mu_3$ and $\Lambda\equiv\frac{\Lambda_2}{2\sqrt{2}}$.
\\\\
Comparing (\refeq{eq:eq42}) with (\refeq{eq:eq34})$\sim$(\refeq{eq:eq36}) and (\refeq{eq:eq44}) with (\refeq{eq:eq16})$\sim$(\refeq{eq:eq18}), they 
perfectly 
agree except for $\sum_im^{2k}_i$.
The first term of eq(\refeq{eq:eq44}):$\frac{\Lambda^2}{16}$ is known to be required from $\text{U}(1)$ factor, but this is beyond the scope of our study and will not be discussed further here.

As we can see above,  
we need to compute to the 8th order term in the instanton expansion 
in order to derive all terms of our 4th order $\mathcal{F}_4$ in our method.
This is because Nekrasov's instanton partition function is a $\Lambda$-expansion
while our expansion is in terms of $a$, where in the latter the coefficients are 
homogeneous polynomials of $\Lambda$ and $m_i$'s, as is always the case 
in such B-model-like approaches.

\section{Conclusions}
\label{Conclusions}
To summarize, we have confirmed that %999
for both cases of $N_{f}=2$ and $3$, $q$ allows an expansion in terms of  
$A\equiv\frac{\Lambda}{8\sqrt{2}a}$  $(N_f=2)$ 
or $\frac{\Lambda}{64\sqrt{2}a}$ $(N_f=3)$ in which the coefficients are 
integer-coefficient polynomials of the moonshine coefficients. As we mentioned 
at the beginning of this paper, this fact suggests an unknown relationship between 
the vertex operator algebra CFT and the Liouville CFT.
The details of that relationship would be worth studying and 
will be discussed elsewhere.  

On the other hand, the method to compute the instanton expansion of the SW prepotential 
developed in \cite{SWMoonshine} and this paper is efficient and simple. We have 
not only re-derived all the known expansions of the $N_f=0,\ldots,3$ prepotentials 
but found some typos in the literature.
For instance, in the $N_f=3$ prepotential presented in section 6 of \cite{CDH}:
(1)   $-a^2(m_1^2 m_2^2 + \cdots)$ in $[\cdots]$ of ${\cal F}^{(2)}$ should read 
$-3a^2(m_1^2 m_2^2 + \cdots)$. (2) In $\{\cdots \}$ of ${\cal F}^{(5)}$, the term 
$-6600 a^6 m_1^2 m_2^2 m_3^2$ is missing. (3) $330(m_1^4m_2^2+\cdots)$ in 
$\{\cdots \}$ of ${\cal F}^{(6)}$ should read $3300(m_1^4m_2^2+\cdots)$.
(4) Also in ${\cal F}^{(6)}$, $2310(m_1^2 m_2^2+\cdots)$ 
should be $2310(m_1^6 m_2^6+\cdots)$. It is also easy and straight-forward to 
derive higher order terms by our method, and will also be applied to E-string theory.

%%% v3 %%%
There are several possible directions to extend the present study.
One direction is to consider a theory with $SU(3)$ or a higher rank gauge group  instead of $SU(2)$. 
This would generally lead to a high genus SW curve, 
which would not fit with the elliptic functions involving the Monstrous moonshine. However, 
 it should be mentioned that ref.\cite{Elias2010.06598} considers
special loci of $SU(3)$ Coulomb branch that parametrize a family of elliptic curves. 

The Hauptmoduln for various rational elliptic surfaces (see e.g.\cite{Horia2203.03755}) 
could also be similarly used as an analogue of the $j$-function to compute prepotentials 
of the corresponding SW theories. For example, the $N_f=2$ theory we considered in this paper 
has the congruence subgroup $\Gamma(2)$ as the monodromy group, whose corresponding 
Hauptmodul is known to be the modular $\lambda$-function. 
Thus, instead of the $j$-function, we could use it to expand $q$ or the prepotential in some variable, 
in which the coefficients would now be polynomials of the coefficients of the McKay-Thompson series, 
and hence of the characters of the Monster.

Finally, let us comment on other moonshine phenomena 
\cite{MathieuMoonshine,UmbralMoonshine}
than the Monstrous moonshine. 
Since we used the special properties of the modular $j$-function to compute 
the instanton partition functions, we were naturally led to think about the Monstrous moonshine. 
On the other hand, in other moonshine phenomena, the Mathieu moonshine for example, 
the expansion of the elliptic genus of K3 in terms of ${\cal N}=4$ superconformal characters,
or equivalently the $q$-expansion of a certain mock modular form,  
gives coefficients which are related to dimensions of representations of $M_{24}$, 
the Mathieu group. 
As already mentioned in Introduction, this connection between the mock modular form 
and the (twisted) SUSY gauge theory is very different from that between the $j$-function and 
SUSY gauge theories in our discussion.
%and it is not clear in what theory such mock modular forms  
%play any role in computing the partition function 
%and other related quantities in the way the $j$-function does in this paper.  
%It should, however, also be mentioned that the elliptic genus of K3 is 
%a Jacobi form of a basic kind, and the elliptic genera of E-strings have been known 
%\cite{Sakai1111.3967,Sakai1706.04619,Sakai2201.06895} to be expressed 
%in Weyl-invariant Jacobi forms. 
%

%%% v3 %%%

\vskip 5mm

We thank Kazuhiro Sakai for valuable discussions on the Mathieu moonshine and E-string theory. 
The authors also thank K.~Ishiguro and R.~Kuramochi for discussions.

\end{document}